\documentclass[prl,twocolumn,aps,showpacs,amsmath,amssymb]{revtex4-1}

\usepackage{graphicx}
\usepackage{epstopdf}
\usepackage{dcolumn}
\usepackage{bm}
\usepackage{amsmath}
\usepackage{graphicx} 
\usepackage{dcolumn}
\usepackage{bm}
\usepackage{amsmath}
\usepackage{epsfig}
\usepackage{float}
\usepackage{epstopdf}
\begin{document}

\title{Non-equilibrium Pair Breaking in Ba(Fe$_{1-x}$Co$_{x}$)$_2$As$_2$ Superconductors: Evidence for Formation of Photo-Induced Excitonic Spin-Density-Wave State}

\author
{X. Yang$^{1}$, L. Luo$^{1}$, M.~Mootz$^{2}$, A. Patz$^{1}$, S. L. Bud'ko$^{1}$, P. C. Canfield$^{1}$, I.~E.~Perakis$^{2}$, and J. Wang$^{1}$}

\affiliation{$^1$Department of Physics and Astronomy and Ames Laboratory-U.S. DOE, Iowa State University, Ames, Iowa 50011, USA. 
	\\$^2$Department of Physics, University of Alabama at Birmingham, Birmingham, AL 35294-1170, USA.}

\date{\today}

\begin{abstract}
Ultrafast terahertz (THz) pump--probe spectroscopy  reveals unusual out-of-equilibrium Cooper pair dynamics  driven by femtosecond (fs) optical quench of superconductivity (SC) in iron pnictides. 
We observe a two--step quench of the SC gap, where an abnormally slow (many 100's of ps) quench process is clearly distinguished from the usual fast (sub-ps)  hot--phonon--mediated scattering channel.   
This pair breaking dynamics depends strongly on  doping, pump fluence, and temperature.
The above observations, together with  quantum kinetic modeling of  non-equilibrium SC and magnetic correlations, provide evidence for photogeneration of a transient  state where SC competes with  
 build--up of spin-density-wave (SDW) excitonic correlation  between  quasi-particles (QP).
\end{abstract}

\pacs{74.25.Gz, 74.20.-z, 78.47.J-, 74.70.Xa, 74.40.Gh}
\maketitle
Ultrafast optical tailoring of transient  quantum states provides a new way 
to discover, design, and control exotic correlated materials phases. 
Recent examples
include, among others, quantum femtosecond magnetism~\cite{Li2013} 
and laser-induced superconductivity~\cite{Cavalleri}.  
This strategic approach is implemented by 
non--thermal separation, within a certain time window, 
of distinct coupled orders. The latter are strongly intertwined in equilibrium, but  respond 
differently  to strong fs  photoexcitation~\cite{Patz_ncomm2014, Porer2014}. 
Iron-arsenide based superconductors (FeSCs) \cite{Kamihara_JACS08} are well--suited for such non--equilibrium 
control, as their properties  are 
determined by  competing SC, SDW, nematic and structural orders   
~\cite{CanfieldARCMP10}.	
Here we address two open issues in FeSCs: (i) how to use non-equilibrium SC pairing/pair breaking to distinguish between two different bosonic  channels, i.e., phonon and SDW, that determine the SC and excitation properties, (ii) how instabilities in two different correlation channels, i.e., Cooper and excitonic, can lead to controllable  transient states.

Ultrafast THz spectroscopy is well--suited for disentangling strongly--coupled excitations. In SCs, this can be achieved  by directly probing out-of-equilibrium Cooper pairs and their dynamics
following strong pump photoexcitation.
By tuning the THz probe frequency in the vicinity of the SC gaps 2$\Delta_{SC}$ of few meV, low-frequency THz electrodynamics can be used to directly measure the time evolution of a SC condensate.   
The latter is ``suddenly" driven away from equilibrium, by fs optical excitation here, as illustrated in Fig.~1(a).  Previous pump--probe experiments showed that the dynamic evolution of a SC condensate following high-frequency optical pump mostly comes from its interactions with hot bosonic excitations \cite{Kabanov2005PRL, Giannetti}. In  FeSCs, both phonon and SDW channels with distinct ultrafast responses are expected to play an important role.
Although the SC dynamics in FeSCs has been measured before in the optical high frequency region \cite{Merteljprl2009}, time-resolved THz spectroscopy experiments in the SC states have been scarce so far.

Photogeneration of non-equilibrium states in quantum materials with competing SC and density wave orders, such as the FeSCs,  provides an opportunity to elucidate the role  of 
electron-hole ($e$-$h$) channels, in addtional to Cooper channel, 
in high--T$_c$ superconductivity. 
 In equilibrium, the SDW phase of FeSCs shows a spontaneous coherence emerging from nested $e$-like and $h$-like Fermi sea pockets, with transition to a (0, $\pi$)/($\pi$, 0) spin-striped state \cite{28NiPRB08,23NandiPRL10}.
Following photoexcitation of QPs in the $e$ and $h$ pockets, an excitonic instability can be triggered by the residual inter--pocket magnetic  interaction 
(illustrated in Fig. 1(b)). 
We refer to  a QP $e$--$h$ pair state as an  excitonic SDW state. Such out-of-equilibrium collective behavior remains, however, elusive so far. 

In this letter, we present an ultrafast THz spectroscopy investigation of the non-equilibrium dynamics of the SC order in Ba(Fe$_{1-x}$Co$_{x}$)$_2$As$_2$
We find that Cooper pair breaking subsequent to strong fs optical excitation follows an unusual two-step temporal profile. In particular, the usual phonon scattering channel ($\tau_{Fast}$) 
is distinguished 
from an additional very slow SC quench ($\tau_{Slow}$). The latter lasts for many 100's of ps under strong pumping. The pump fluence dependence of the SC quench  
differs significantly between the under- and overdoped regimes with different SDW coherence. 
The remarkably slow pair--breaking dynamics, together with quantum kinetic modeling, provide evidence for the formation of a non-equilibrium correlated state of QP $e$--$h$ pairs competing with SC, which is driven by excitonic correlation of the disconnected Fermi sea pockets. 

The samples are single-crystalline Ba(Fe$_{1-x}$Co$_{x}$)$_2$As$_2$ with cobalt substitutions of $x$=0.047 and 0.1. 
In the underdoped sample ($x$=0.047), long-range SDW and structural phase transitions appear at T$_\text{N}$=48~K and T$_\text{S}$ =66~K, respectively~\cite{23NandiPRL10}. The phase transitions are {\em absent} in the overdoped sample ($x$=0.100). 
Both samples exhibit a SC transition at T$_C \sim$ 17~K. Our optical pump--THz  probe reflectivity spectroscopy setup 
is described in detail elsewhere~\cite{luo2015}. 
The opaque sample is mounted at 45$^{o}$ to incident light and cooled to T=4.1~K.

\begin{figure}[!tbp]
	\begin{center}
		\includegraphics[scale=.37]{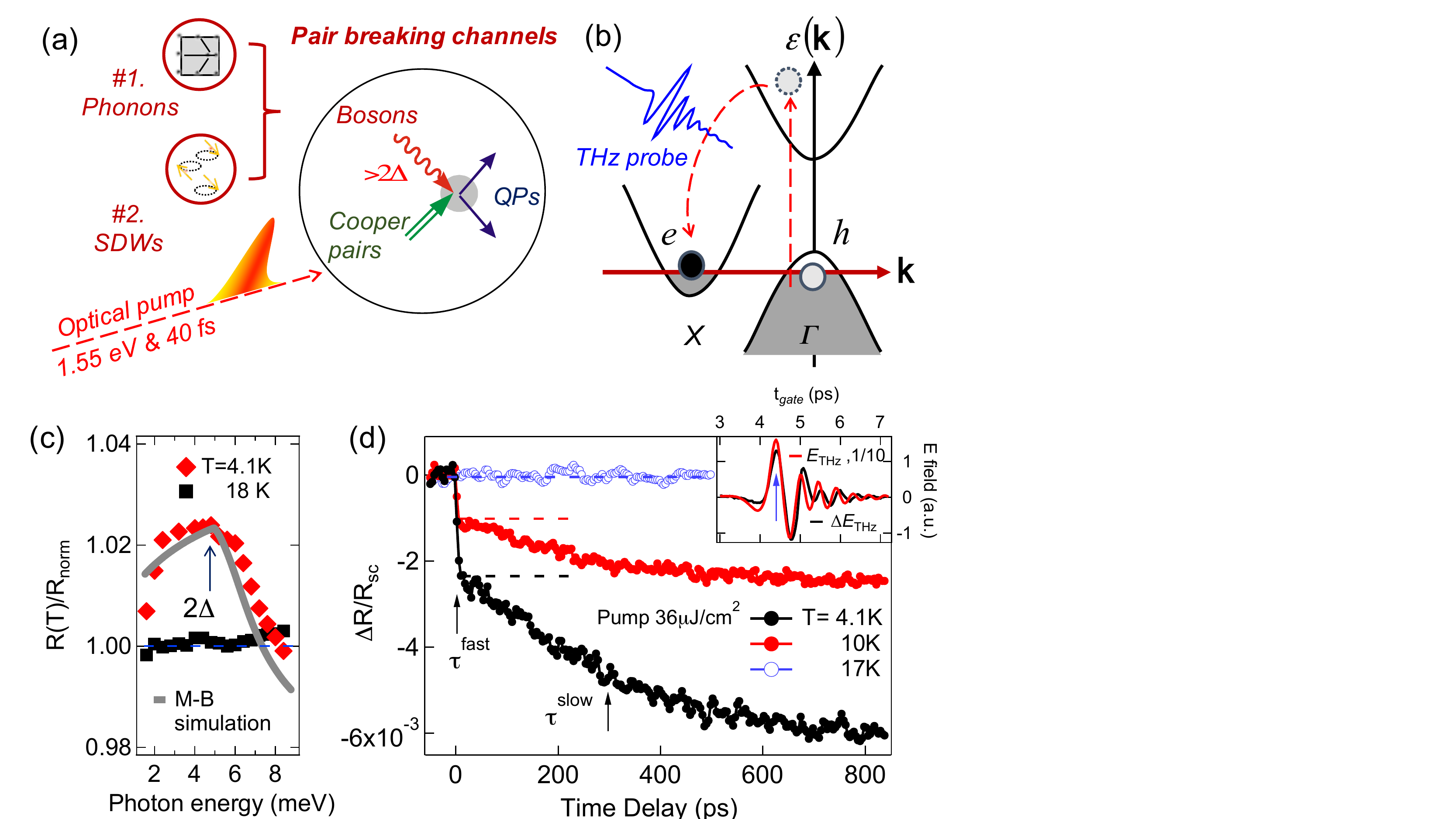}
		\caption{Schematics of SC pair breaking channels, (a), and interband transitions, (b), after fs pump photoexcitation.  (c): Static THz reflectivity spectra, normalized to the normal state spectra at 20~K, for underdoped $x=0.047$ sample, at 4.1~K and 18~K. Grey line shows the result of the Mattis-Bardeen theory. (d) Ultrafast THz dynamics for the above underdoped sample. Inset: The measured time--dependent THz field transients, with gate-time (blue arrow) t$_{gate}$=4.4~ps, at $T$=4.1K.}
		\label{fig1} 
	\end{center}
\end{figure}	

We start with the equilibrium THz measurements of the static SC order and energy gap. The typical static THz reflection spectra, $R(T)$, of Ba(Fe$_{0.953}$Co$_{0.047}$)$_2$As$_2$  are shown in Fig.~1(c). We compare temperatures $T$=4.1~K  (red diamonds) and 18~K (black rectangles), below and above the SC transition respectively. 
These spectra are obtained through  Fourier transform of the measured time domain THz field traces, e.g., the red-line curve in the inset of Fig.~1(d). They are normalized by the normal state 20~K trace (not shown). 
The ratio $R(4.1\,K)/R(20\,K)$ in the measured spectral range of 1--11~meV exhibits the characteristic SC profile. The distinct upward cusp with maximum at $\sim$5~meV reflects the SC energy gap 2$\Delta_{SC}$. 
In contrast,  $R(18\,K)/R(20\,K) \sim 1$ has a featureless spectral shape. The measured reflectivity spectra are reproduced well by the Mattis-Bardeen (MB) theory  and by Fresnel equations.
In the low-frequency/temperature limit, the measured ratio can be expressed as $1+4\sqrt{\omega /(\pi \sigma _{1N})}$, where $\sigma _{1N}$ is the normal state conductivity.

The ultrafast THz differential reflectivity $\Delta R/R_{SC}$ in the underdoped compound is shown in Fig.~1(d) for three different temperatures, 4.1~K, 10~K and 17~K.  The pump fluence and photon energy are set to 40~$\mu$J$/cm^{2}$ and 1.55~eV, respectively. The transient signals are 
given by the difference of the time-dependent THz fields in the photo-excited (pump on, back line, inset) and unexcited (pump off) states (red line, inset). $\Delta R/R_{SC}$ is then obtained as $[(E_{THz}+\Delta E_{THz})^{2}-E{_{THz}^{2}}]/E{_{THz}^{2}}$.  The $\Delta R/R_{SC}$ dynamics at a fixed gate time $t_{gate}= 4.4$~ps (blue arrow, inset)  is recorded as function of pump--probe delay. Fig.~1(d) demonstrates  a distinct two-step temporal profile of pair--breaking dynamics. The initial sub-ps SC gap  decrease ($\tau_{\mathrm{Fast}}$) is followed by a further very slow SC quench that lasts for an unusually long time $\sim$800~ps ($\tau_{\mathrm{Slow}}$). The strong temperature dependence in Fig.~1(d) coincides with the SC transition. Approaching the critical temperature from below, the transient signals quickly decrease, as seen in the 4.1~K (black circle) and 10~K traces (red), and diminish at T$\approx$T$_C$ $\approx$ 17~K (blue).  
\begin{figure}[!tbp]
	\begin{center}
		\includegraphics[scale=.4]{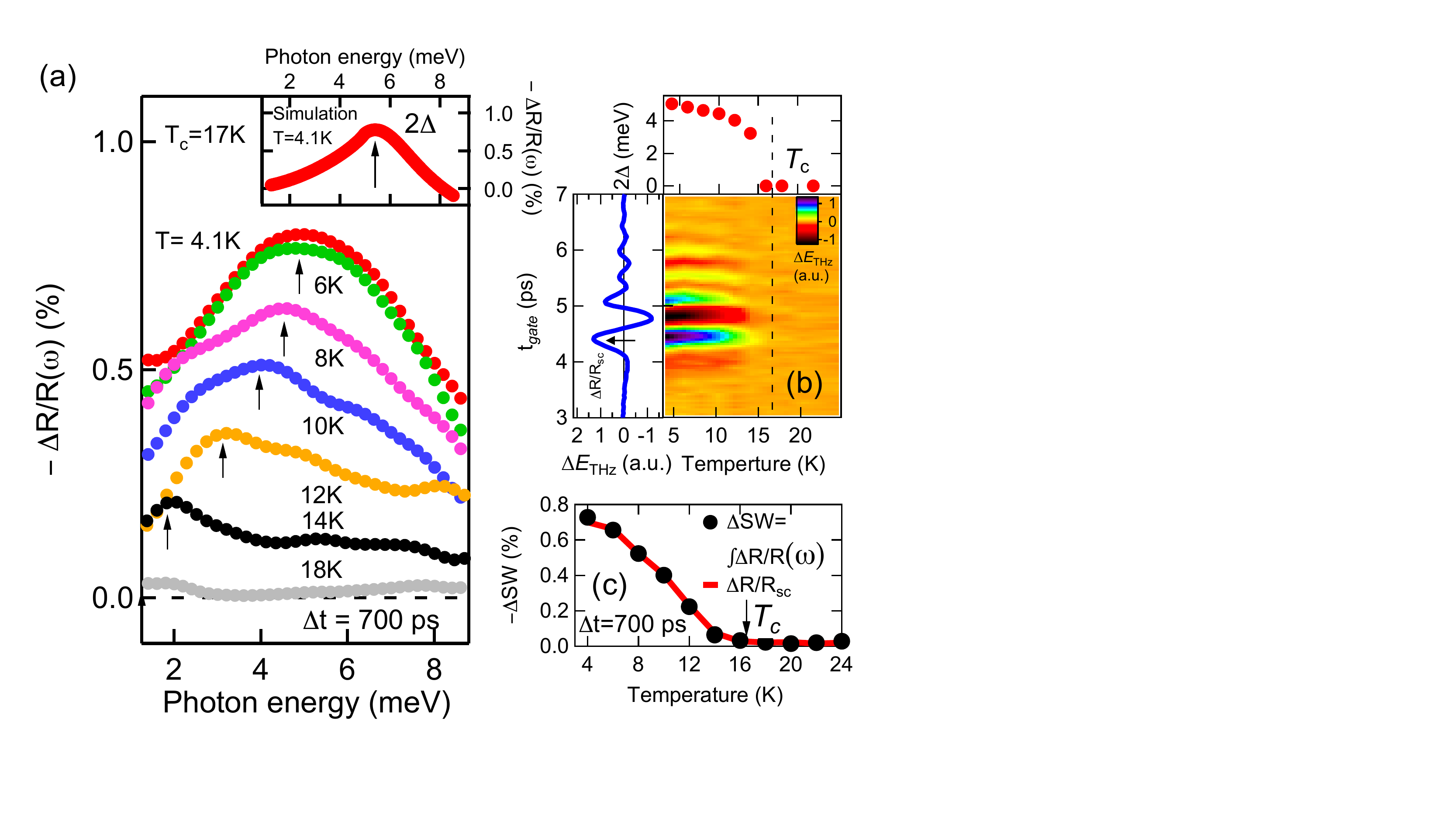}
		\caption{(a) THz differential reflectivity spectra (dots) for the $x$=0.047 sample at 700~ps. The cusp peak marked by black arrows reflects 2$\Delta_{SC}$.
			Inset shows the MB simulation (see text). (b) 
			Temperature--dependent $\Delta$E/E THz transients. Left panel: $\Delta$E/E transient at 4.1~K. Top panel: temperature dependence of 2$\Delta_{SC}$.
			(d) Temperature dependence of the integrated spectral weight and peak transient amplitude.}
		\label{fig2} 
	\end{center}
\end{figure}	

We now present transient $\Delta R(\omega)/R_{SC}$ spectra that further point to non-equilibrium pair--breaking as the origin of the pump-induced THz signals.  Figure~2(a) shows the temperature--dependent low frequency,
$\sim$1--9~meV, 
differential reflectivity spectra of the underdoped $x=0.047$ sample at a fixed long time delay of 700~ps.
These spectra are obtained from the Fourier transform of the time-domain THz raw data (Fig.~2(b)). We note three distinct features of $\Delta R(\omega)/R_{SC}$: (1) The negative low frequency change $\Delta R(\omega)/R_{SC}<$0 indicates photo-induced condensate breaking processes. (2) The pump-probe transient spectra exhibit the characteristic SC lineshape with cusp peak at 2$\Delta_{SC}$ (black arrow). This lineshape can be reproduced well by the MB theory, as shown in the inset of Fig.~2(a).  
(3) Approaching the critical temperature from below, we see that  $\Delta R(\omega)/R_{SC}$ quickly diminish as  the cusp at 2$\Delta_{SC}$ shifts to lower frequencies (black arrows, Fig.~2(a)). This SC gap temperature dependence is summarized in the top panel of Fig.~2(b). 
In addition, we compared  the integrated reflectivity spectral weight associated with the SC states and the $\Delta R/R_{SC}$ amplitude at t$_{gate}$=4.4~ps. The strong correlation between the two at all temperatures (Fig.~2(c)) and fluences (inset, Fig.~4(b)), allows us to study the ultrafast pair-breaking dynamics by recording $\Delta R/R_{SC}$ as in Fig.~1(d).

\begin{figure}[!tbp]
	\begin{center}
		\includegraphics[scale=.32]{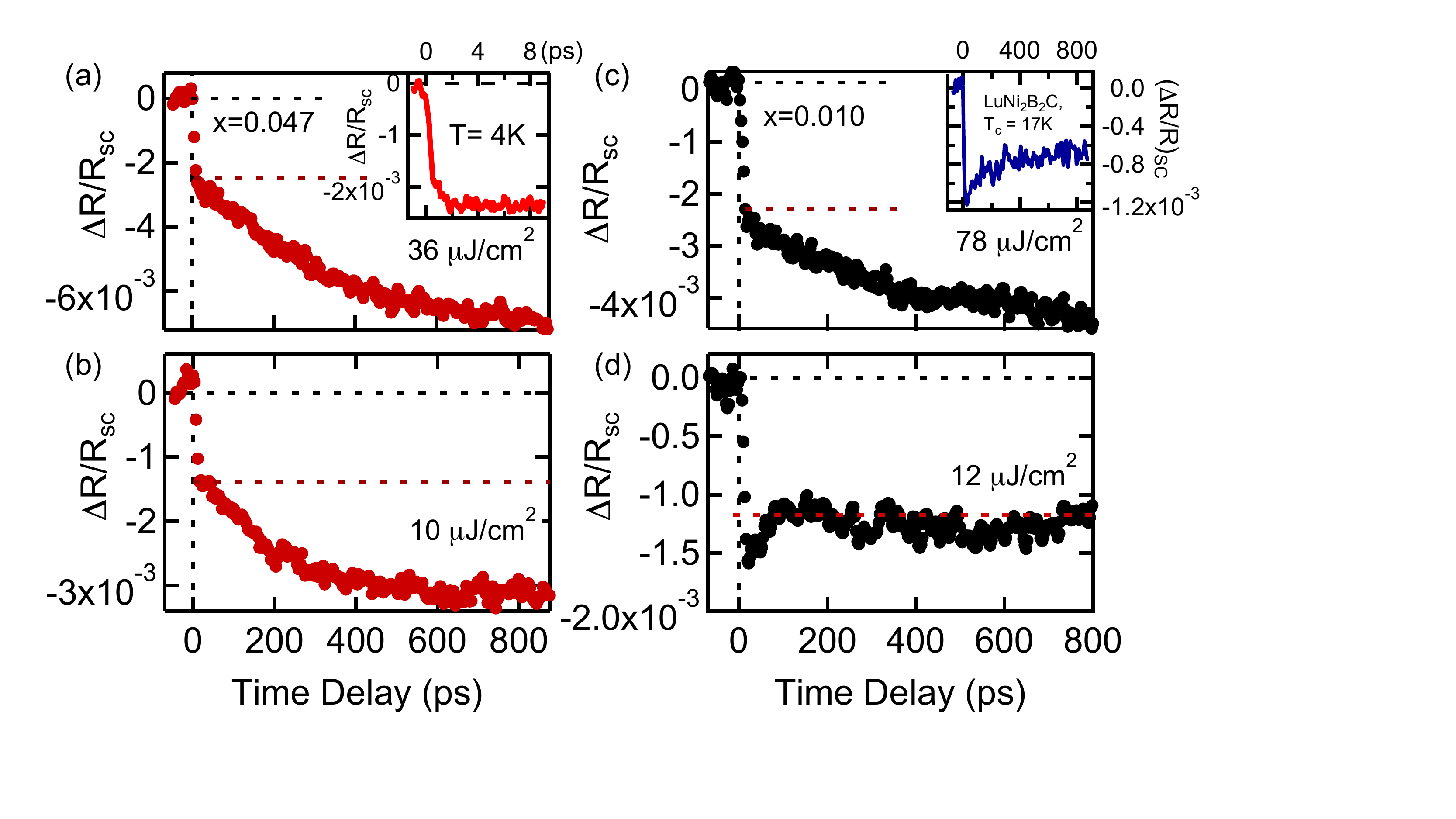}
		\caption{Ultrafast THz pump probe scan at different pump fluences for (a,b) $x$=0.047 and (c,d) $x$=0.1 samples. All traces taken in the superconducting state at T=4.1~K. Inset of (a): the initial dynamics. Inset of (c): The THz dynamics in LuNi$_2$B$_2$C at pump fluence of 40~$\mu$J/cm$^{2}$.}
		\label{fig3} 
	\end{center}
\end{figure}

Next we show the strong dependence of the non-equilibrium SC quench profile on pump fluence and doping. 
Figs. 3(a) and 3(b) show the photoinduced $\Delta R/R_{SC}$ dynamics in the underdoped, $x$=0.047, sample and compare  36~$\mu J/cm^{2}$ and 10~$\mu J/cm^{2}$ pumping. Both excitations of the coupled SC/SDW ground state
order show a sub-ps $\tau_{\mathrm{Fast}}$ followed by a 100's ps $\tau_{\mathrm{Slow}}$ process. Previous works in BCS and cuprate SCs have shown that the majority of the absorbed photon energy transfers to the phonon reservoir during the pulse \cite{Giannetti}. Hot phonons then deplete the condensate on a few-ps timescale \cite{Beck2011PRL, Kabanov2005PRL}. This time interval becomes shorter (sub-ps) under the strong pumping used here, consistent with the the inset of Fig.~3(a). Here the slow $\sim$800ps SC quench under strong pumping appears to be different from other superconductors (supplementary). For comparison, the inset of Fig.~3(c) shows 
the non-equilibrium pair breaking dynamics of the BCS superconductor LuNi$_2$B$_2$C.
Unlike for the FeSCs, similar strong pumping of this superconductor  exhibits single-step, sub-ps SC quenching, followed by \textit{partial recovery} and long decay. 
This typical pair breaking temporal profile is consistent with the $\tau_{\mathrm{Fast}}$ component in the FeSCs. It can be explained in terms of QP scattering with high energy phonons, followed by condensate recovery governed by phonon relaxation \cite{Kabanov2005PRL}. 
A comparison of the two SC systems indicates that the additional, remarkably slow and yet strong, channel is exclusively present in the FeSCs. 
This appears to be distinct from the usual  hot phonon bosonic channel.
The continuing SC gap quench over many 100’s of ps is ``intrinsic" and unlikely to come from, e.g., heat diffusion, which would appear in both studied systems.  

Figs.~3(c) and (d) show our results 
in the overdoped FeSC system ($x$=0.1),
where there is no long-range SDW order in equilibrium.
In this regime 
of the phase diagram, 
the quench temporal profile  
changes drastically with increasing pump fluence different from the underdoped regime. For example, while the slow SC quench is again seen at high fluences 78~$\mu\mathrm{J}/\mathrm{cm}^2$ (Fig.~3(c)), at low pump fluences (12~$\mu\mathrm{J}/\mathrm{cm}^2$ in Fig.~3(d)) the initial fast quench is followed by a partial recovery similar to the BCS sample (inset, Fig. 3(c)).  
Our results show  that, for the overdoped ground state without SDW coherence, 
the slow SC quench channel only appears above a critical fluence. While in the underdoped regime with SC/SDW ground state, it persists down to much lower fluences.  
Such a strong distinction between sample doping corroborates that the continuing SC gap quench is ``intrinsic" in FeSCs that differs from both BCS and cuprate SCs.

\begin{figure}[!tbp]
	\begin{center}
		\includegraphics[scale=.42]{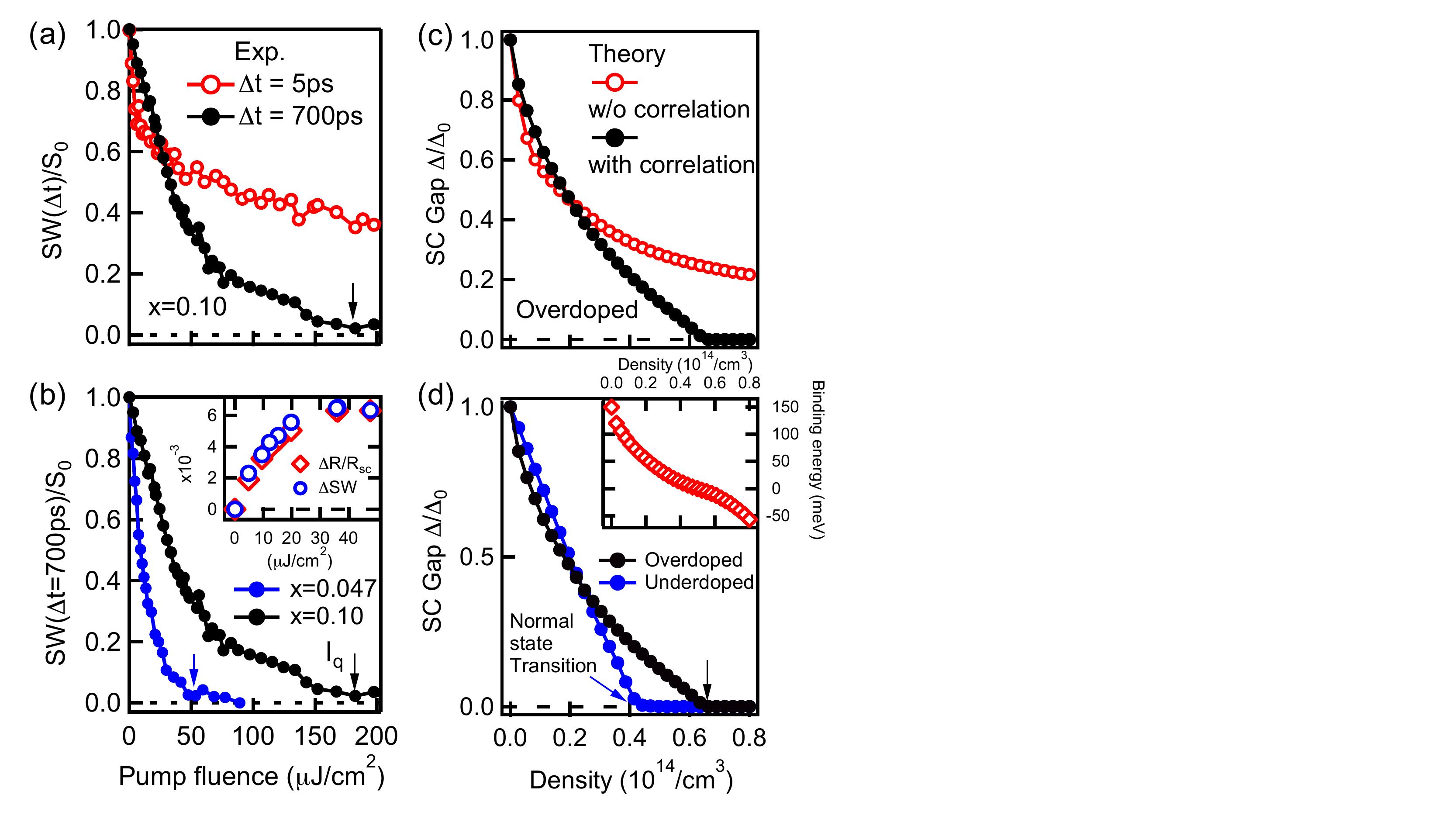}
		\caption{Measured fluence dependence of the integrated spectral weight (SW): (a) $x$=0.1 crystal at 5~ps (red) and 700~ps (black); (b) comparison of $x$=0.047 and $x$=0.1 samples at 700~ps. Inset: SW and $\Delta$R/R$_{SC}$ exhibit the same fluence dependence. (c) Theoretical modeling of the SC gap quench in the overdoped region  as function of photoexcited QP density $\rho$,  with (black line) or without (red line)   inter-pocket excitonic correlation.
The $y$-axis is normalized by the equilibrium SC gap  $\Delta_0$. 
			(d) Theoretical comparison of under- and over-doped regions for the photoinduced correlated SDW excitonic state. Inset: Excitonic  energy $|E|$, 
Eq. (\ref{eq:Wannier0}), as function of  $\rho$.}
		\label{fig4} 
	\end{center}
\end{figure}

The striking  fluence and doping dependences of the FeSC condensate quench  are seen more clearly  in Figs.~4(a) and 4(b). Here, the integrated spectral weight (SW), obtained from the peak-peak amplitude change $\Delta R/R_{SC}$ at t$_{gate}$=4.4~ps (inset, Fig.~4(b)), is shown as function of pump fluence and compared between underdoped and overdoped regimes. 
Fig.~4(a) compares the fluence dependence 
in the overdoped regime (no SDW ground state coherence) between 
short 5~ps (red empty circle) and long 700~ps (black solid circle) time delays. The  SC quench 
as function of photocarrier density 
is qualitatively different at  short and long  times and the two curves cross at $\sim$25~$\mu\mathrm{J}/\mathrm{cm}^2$.
In particular, at  $\Delta t$=700~ps, we observe a transition from SC to normal state above a large critical pump fluence, I$q$=182~$\mu\mathrm{J}/\mathrm{cm}^2$. Such transition is not observed at 5ps in the measured range, where the 
signal appears to saturate for high fluences.  
Fig.~4(b) compares this  SC-to-normal state transition at  $\Delta t$=700ps  
between the under- and overdoped samples. 
In the underdoped regime 
with SC/SDW coherence
($x$=0.047), the transition 
 occurs
at much smaller critical pump fluence $\sim$ 50~$\mu\mathrm{J}/\mathrm{cm}^2$ than in the overdoped regime without SDW coherence.    
 Below we provide an interpretation of these  salient experimental features based on  
quantum kinetic calculations of
photoinduced 
 build--up of SDW excitonic correlation  between the photoexcited $e$ and $h$ QPs.

The conventional  Rotwarth-Taylor (RT) model
 \cite{Kabanov2005PRL, RT}, which describes  
QP interactions with a hot boson bath 
whose properties are not strongly affected 
by photoexcitation, 
does not provide a consistent fit of our experimental data (supplementary). 
There 
we summarize a density matrix 
equation--of--motion  
calculation of the QP population dynamics
arising from the strong inter--pocket magnetic interaction in FeSCs. 
We focus on the build--up  of 
$e$--$h$ excitonic correlation 
after  initial   photocarrier  relaxation ~\cite{Porer2014}. The latter results  in hot QP 
populations of  disconnected $e$-- and $h$--like Fermi sea pockets (Fig. 1(b)). Subsequently, interband residual 
interactions between such  QPs can  
form    excitonic pairs prior to inter--pocket relaxation that
requires a large momentum transfer (Fig. 1(b)). Slow ps formation of Coulomb-bound excitonic correlation prior to interband recombination has been previously established  in semiconductors  \cite{Book:11}.

Our starting point for describing the 
QP population dynamics is the model
of  Refs.~\cite{Vorontsov:2010,Fernandes:2010}.
This model captures the most essential 
features of the competition between SC and SDW orders in the FeSCs.
Assuming for simplicity 
one $e$  and one $h$ pocket,  
we first transform to the basis of  Bogoliubov quasi-particles with both SC and SDW coherence.
We then study the effects of the residual 
inter--pocket interaction among these QPs in the SDW channel. 
High frequency pump excitation  and  fast photocarrier scattering with phonons creates QP  incoherent distributions  in both $e$ and $h$ pockets within 100fs timescales \cite{Porer2014}.
 These hot  distributions create an 
initial condition for subsequent non--thermal time evolution  
of  QP $e$ and $h$ populations $n_{{\bf p}}$ coupled by inter--pocket $e$--$h$ 
correlation.  
The latter is driven  by the residual QP interband interaction, which 
leads to Coulomb--correlated QP pairs (spin excitons) with total momentum 
determined  by the difference between the $e$ and $h$ pockets.   
For our purposes here,  
we  consider quasi--stationary solutions of the equations of motion.
Prior to inter-pocket relaxation, 
we thus obtain a 
correlated SDW  state 
formed by  QP $e$--$h$ pairs with   
amplitude $\phi_p$. The latter is   
described by  the 
generalized Wannier equation (supplementary) 
\begin{align}
\label{eq:Wannier0}	\left(\varepsilon^{-}_\mathbf{p}+
\varepsilon^{+}_\mathbf{p}\right)\phi_\mathbf{p}-\left(1-2\,n_\mathbf{p}\right)\sum_\mathbf{k}V_{\mathbf{k},\mathbf{p}}\phi_\mathbf{k}=E\,\phi_\mathbf{p}\,.
\end{align}
Here, $\varepsilon^{-}_\mathbf{p}$ and 
$\varepsilon^{+}_\mathbf{p}$ are the 
single--QP energies 
and $V_{\mathbf{k},\mathbf{p}}$ are the 
interband Coulomb matrix elements describing the residual interaction.  
The energy eigenvalue $E$ describes excitonic corrections to the chemical potential, which develop as correlation builds up between the photoexcited QPs.
$E$ depends on the total photoexcited QP
density 
$\rho$ through the condition
$\rho=\sum_{{\bf p}} n_{{\bf p}}$, where the QP distribution
$n_\mathbf{p}$ is 
determined by (supplementary)  
\begin{align}
\label{eq:conserv}
\left(n_\mathbf{p}-\frac{1}{2}\right)^2+|\phi_\mathbf{p}|^2=\frac{1}{4}.
\end{align} 
The above equations, together with 
SC/SDW order parameter 
equations discussed in the supplement, 
provide a self-consistent 
calculation of a correlated 
SDW/SC photoexcited  state 
prior to inter--pocket  recombination. 
They are  similar to the  description of incoherent excitonic correlation build--up in semiconductors ~\cite{Book:11}  and transition between 
Bose condensation and 
BCS superconductivity in the case of fermions with attractive 
interaction~\cite{Nozieres}. 
Here we describe QP $e$--$h$ pair states 
whose properties  depend on the SC and SDW  coherence
analogous to  Ref.~\cite{Bardasis}. This  
introduces a doping dependence that depends on the Fermi sea topology. 
Eq. ~(\ref{eq:Wannier0}) interpolates between the weak and strong coupling limits and can have both bound and unbound solutions,
depending on  QP density $\rho$, Pauli blocking effects, 
and inter--pocket interaction 
strength.
With increasing $\rho$, the momentum 
dependence and magnitude of the QP distribution $n_{{\bf p}}$ changes strongly due to 
its coupling with the QP pair amplitude $\phi_{{\bf p}}$. This, in   turn, affects the Pauli blocking effects   that quench  the SC gap.

Fig.~4(c) compares the calculated quasi-stationary SC gap with or without the excitonic correlation  $\phi_{\bf p}$ as a function of 
total QP density $\rho$. 
Since the pump--induced photocarriers  have already relaxed during the initial ps quench, 
we model the initial condition by assuming quasi-equilibrium Fermi--Dirac QP distributions
with hot temperature determined by the total 
$\rho$ (supplementary material).
Fig.~4(c) shows the effects of  SDW  
exciton formation on the SC gap   in the overdoped regime, where only SC order is present in the ground state. 
The calculated SC gap without excitonic  QP correlation, $\phi_p$=0 (red circles), 
shows a fast decrease  at low QP densities, which  flattens (saturates) as $\rho$ increases further.
This feature is  
in qualitative agreement with the measured fluence dependence of the SC gap at short ps time delays (red circles, Fig.~4(a)).
During short timescales, excitonic correlation has not had time to built--up yet and thus $\phi_p \approx$ 0. 
The Pauli blocking saturation behavior of the SC gap with increasing $\rho$ results from the momentum dependence of $n_{{\bf p}}$, which for $\phi_{{\bf p}}$=0 is similar to the simple BCS theory~\cite{Axt}.

With $\phi_p \ne$0  at later times, 
the QP distribution $n_{{\bf p}}$  changes
drastically. Our calculations show that
 SDW excitonic formation leads to a complete quench of the SC gap at elevated $\rho$
(black curve in Fig. 4(c)). 
The complex photoinduced interplay of an emergent correlated SDW state with the SC condensate
thus results in a  transition from SC  to normal state above a critical pump fluence. 
This result  is in qualitative agreement  with the measured fluence dependence of the SC gap at 700\,~ps (black circles, Fig.~4(a)). It   
supports our claim that the experimentally observed qualitative 
difference  in the SC gap fluence dependence between  long and short times 
arises from the delayed formation of a correlated SDW state. The latter requires build--up of excitonic 
correlation between the laser-induced $e$ and $h$ QP populations.

Figure~4(d) compares the calculated QP density dependence of the SC gap at long times between  the underdoped (blue solid circle) and overdoped  (black solid circle) regimes.
The  doping dependence of the critical pump fluence required 
for SC--to--normal state transition  is 
in qualitative agreement with the experiment (compare to Fig. 4(b)). 
While in the overdoped regime SC order can only compete with photoinduced  SDW excitonic order,   the additional  condensed SDW  coherence present in the underdoped regime results  in  SC-to-normal state transition at lower photoexcited QP densities. 
Such doping dependence of the residual $e$-$h$ correlations comes from the 
differences in coherence  and Fermi sea pockets  between the overdoped and underdoped regimes.     
Finally, the inset of Figure~4(d) shows the calculated  energy eigenvalue $E$ per $e$--$h$ pair as a function of  QP density  $\rho$. At low $\rho$, we obtain a bound SDW excitonic state, which however becomes unbound  due to phase-space filling Pauli effects. 

In conclusion, we performed the first ultrafast THz spectroscopy in FeSCs to demonstrate
the existence of an additional, remarkably slow, pair--breaking channel. 
This is consistent with our theoretical prediction that the build-up of SDW excitonic correlation between laser-induced $e$ and $h$ QPs manifests itself in the long--time THz formation dynamics with strong pump--fluence and doping dependence.  
The fast control demonstrated here by adjusting both pump fluence and doping may be used to access hidden density-wave phases and quantum criticality under the SC dome in high--$T_c$ superconductors. 
\section*{Acknowledgement}
This work was supported by the Army Research office under award W911NF-15-1-0135 (THz spectroscopy). Theory work at the University of Alabama, Birmingham was supported by start--up funds. Sample growth, characterization and phase diagram analysis (PCC and SLB) was supported by the Ames Laboratory, the US Department of Energy, Office of Science, Basic Energy Sciences, Materials Science and Engineering Division under contract \#DE-AC02-07CH11358.
THz instrument was supported in part by the Keck Foundation (J.W.).


\section*{APPENDIX}  

\section{Rothwarf-Taylor model} 

In the conventional Rothwarf--Taylor model, the dynamics of QP and hot boson populations, $\rho(t)$ and $N(t)$ respectively, are described by coupled differential equations~\cite{RT}. In the strong bottleneck regime, where the decay rate of the hot bosons is small, the pre-bottleneck dynamics can be solved analytically, yielding the QP dynamics~\cite{Kabanov2005PRL}
\begin{align}
\label{eq:n_t}
\rho(t)=\frac{\beta}{R}\left[-\frac{1}{4}-\frac{1}{2\tau}+\frac{1}{\tau}\frac{1}{1-K\,\mathrm{exp}(-t\beta/\tau)}\right]\,.
\end{align}
Here, $R$ is the QP recombination rate while $\beta$ defines the probability of pair-breaking by hot bosons. The dimensionless parameters 
\begin{align}
\label{eq:RT_para}
K=\frac{\frac{\tau}{2}\left(\frac{4\,R\,\rho_0}{\beta}+1\right)-1}{\frac{\tau}{2}\left(\frac{4\,R\,\rho_0}{\beta}+1\right)+1}\,,\quad \frac{1}{\tau}=\sqrt{\frac{1}{4}+\frac{2R}{\beta}\left(\rho_0+2\,N_0\right)}
\end{align}
are determined by the photoexcited initial  QP ($\rho_0$) and hot boson ($N_0$) populations, with  -1$\le K\le$1
~\cite{Kabanov2005PRL}.
To obtain the parameters $\beta$ and $R$, we fit the time-delay traces of the measured THz differential reflectivity $\Delta R/R_{SC}$ shown in Fig.~3 by using Eq.~(\ref{eq:n_t}) as in Refs.~\cite{Demsar,Beck2011PRL}. However, our  fits produce unphysical $K<$-1. This indicates that the conventional Rothwarf--Taylor model does not provide a consistent description of the  many 100's of ps relaxation observed in FeSCs with increasing strong pump photexcitation. 

\section{Many-body theory of excitonic SDW correlation formation}

\subsection*{Hamiltonian}

In this section we provide a brief overview of our microscopic theory, with  
full details and further calculations to be presented elsewhere \cite{theo_paper}. 
We use the simplest model Hamiltonian
believed to give a  good 
qualitative description of the 
competition between SC and SDW order in the iron pnictides \cite{Fernandes:2010,Vorontsov:2010}:
\begin{align}
\label{eq:Htot}
H=H_0+H_\Delta+H_\mathrm{m}\,.
\end{align}
The non-interacting part of the Hamiltonian describes the $e$ and $h$ Fermi sea pockets
predicted by the bandstructure:  
\begin{align}
H_0=\sum_{\mathbf{k},\sigma}[\xi_\mathrm{c}(\mathbf{k})c^\dagger_{\mathbf{k},\sigma}c_{\mathbf{k},\sigma}+\xi_\mathrm{f}(\mathbf{k})f^\dagger_{\mathbf{k},\sigma}f_{\mathbf{k},\sigma}]\,.
\end{align}
We include only one circular hole-like band at  the $\Gamma$-point, with dispersion $\xi_\mathrm{c}(\mathbf{k})=\xi_\mathrm{c,0}-\frac{\hbar^2 k^2}{2m_\mathrm{c}}-\mu$, and one elliptical electron-like band with dispersion $\xi_\mathrm{f}(\mathbf{k})=\frac{\hbar^2 k_x^2}{2m_{\mathrm{f}_x}}+\frac{\hbar^2 k_y^2}{2m_{\mathrm{f}_y}}-\xi_\mathrm{f,0}-\mu$ close to the $\mathbf{Q}_0=(\pi,0)/(0,\pi)$ pocket \cite{Fernandes:2010,Vorontsov:2010}. These electron and hole energy  dispersions are determined by effective masses $m_{\mathrm{c}/\mathrm{f}}$, energy offsets $\xi_\mathrm{\mathrm{c}/\mathrm{f},0}$, and chemical potential $\mu$. The operators $f^\dagger_{\mathbf{k},\sigma}$ ($c^\dagger_{\mathbf{k},\sigma}$) create a  
carrier with crystal momentum $\hbar(\mathbf{k}-\mathbf{Q}_0)$ ($\hbar\mathbf{k}$) and spin $\sigma$ in the electron-like band near $\mathbf{Q}_0$  (hole-like band close to the $\Gamma$-point).
The SC pairing interaction~\cite{Fernandes:2010}
is given by 
\begin{align}
\label{eq:HSC}
H_\Delta=V_\mathrm{SC}\sum_{\mathbf{k},\mathbf{k}'}\left[c^\dagger_{\mathbf{k},\uparrow}c^\dagger_{-\mathbf{k},\downarrow}f_{-\mathbf{k}',\downarrow}f_{\mathbf{k}',\uparrow}+\mathrm{h.c.}\right]\,,
\end{align}	 
with interaction magnitude $V_\mathrm{SC}$. 
Here we only include the pair hopping between the two pockets~\cite{Fernandes:2010,Vorontsov:2010}, which is 
believed to be the dominant interaction  producing  $s^{+-}$ SC pairing~\cite{Akbari:2013}. Besides this SC interaction, the low energy properties  depend on  the magnetic interaction in the SDW channel~\cite{Fernandes:2010}
\begin{align}
\label{eq:Hm}
&H_\mathrm{m}=-\frac{V_\mathrm{m}}{2}\sum_{\mathbf{k},\mathbf{p},\mathbf{q}}S^\dagger_z(\mathbf{p},\mathbf{q})S_z(\mathbf{k},\mathbf{q})\,,\nonumber \\ &
S_z(\mathbf{k},\mathbf{q})=\sum_{\sigma}\sigma\,c^\dagger_{\mathbf{k},\sigma}f_{\mathbf{k}+\mathbf{q},\sigma}\,,
\end{align}	
where $V_\mathrm{m}$ describes  the strength of the magnetic interaction.
For simplicity we neglect the coupling to phonons.

\subsection*{Ground-state configuration}
Following previous works, the  equilibrium SC properties can be described by treating 
the above Hamiltonian Eq.(\ref{eq:Htot}) in the mean-field approximation  ~\cite{Fernandes:2010,Vorontsov:2010}:
\begin{align}
\label{eq_HSC_MF}
H^{\mathrm{MF}}_\Delta=-\sum_{\mathbf{k}\in\mathcal{W}}\left[\Delta_\mathrm{c}c_{-\mathbf{k},\downarrow}c_{\mathbf{k},\uparrow}+\Delta_\mathrm{f}f_{-\mathbf{k},\downarrow}f_{\mathbf{k},\uparrow}+\mathrm{h.c.}\right]\,,
\end{align}	 
where the SC order parameters $\Delta_\mathrm{c}$ and $\Delta_\mathrm{f}$ are given by ~\cite{Fernandes:2010}
\begin{align}
\label{eq:delta}
&\Delta_\mathrm{c}=-V_\mathrm{SC}\sum_\mathbf{p\in\mathcal{W}}\langle f_{-\mathbf{p},\downarrow}f_{\mathbf{p},\uparrow}\rangle\,, \nonumber \\
&\Delta_\mathrm{f}=-V_\mathrm{SC}\sum_\mathbf{p\in\mathcal{W}}\langle c_{-\mathbf{p},\downarrow}c_{\mathbf{p},\uparrow}\rangle\,.
\end{align}
The sums in Eq.(\ref{eq:delta}) only include the set $\mathcal{W}$ of wavevectors $\mathbf{k}$ with $|\xi_\lambda(\mathbf{k})|\le\hbar\omega_\mathrm{C}$, where $\omega_\mathrm{C}$ is the cut-off frequency. Hartree-Fock decoupling of the magnetic interaction Eq.(\ref{eq:Hm}) gives ~\cite{Fernandes:2010}
\begin{align}
H^\mathrm{MF}_\mathrm{m}=-\sum_{\mathbf{k},\sigma}\sigma\left[M\,f^\dagger_{\mathbf{k}+\mathbf{q},\sigma}c_{\mathbf{k},\sigma}+M\,c^\dagger_{\mathbf{k},\sigma}f_{\mathbf{k}+\mathbf{q},\sigma}\right]
\end{align}
with SDW order parameter ~\cite{Fernandes:2010}
\begin{align}
M=\frac{V_\mathrm{m}}{2}\sum_{\mathbf{k},\sigma}\sigma\, \langle c^\dagger_{\mathbf{k},\sigma}f_{\mathbf{k}+\mathbf{q},\sigma} \rangle\,
\end{align}
assumed to be polarized in the $z$-direction~\cite{Brydon:2009}. The SDW order is assumed for simplicity to form with a single  momentum $\mathbf{Q}=\mathbf{Q}_0+\mathbf{q}$ between the electron and hole pockets, which becomes commensurate for $\mathbf{q}=0$. We only include this momentum in our calculations here. 
The mean-field Hamiltonian $H^\mathrm{MF}=H_0+H^\mathrm{MF}_\mathrm{SC}+H^\mathrm{MF}_\mathrm{m}$ is diagonalized exactly, yielding the self-consistent temperature--dependent SC and SDW gap equations that characterize the 
thermal equilibrium state 
~\cite{Fernandes:2010}:
\begin{align}
\label{eq:gap_gs}
&\Delta_\lambda=-\frac{V_\mathrm{SC}}{S}\sum_{\mathbf{k},j}K^\lambda_{\mathbf{k},j}\mathrm{tanh}\left(\frac{E_{j,\mathbf{k}}}{2\,k_\mathrm{B}T}\right)\,,\nonumber \\
&M=\frac{V_\mathrm{m}}{S}\sum_{\mathbf{k},j}K^\mathrm{m}_{\mathbf{k},j}\mathrm{tanh}\left(\frac{E_{j,\mathbf{k}}}{2\,k_\mathrm{B}T}\right)\,,
\end{align}
with kernels
\begin{align}
&K^\lambda_{\mathbf{k},j}=\frac{\Delta_{\bar{\lambda}}(E^2_{j,\mathbf{k}}-\Delta^2_\lambda-\xi^2_{\lambda,\mathbf{k}})+M^2\Delta_\lambda}{2\,E_{j,\mathbf{k}}(E^2_{j,\mathbf{k}}-E^2_{\bar{j},\mathbf{k}})}\,, \nonumber \\
&K^\mathrm{m}_{\mathbf{k},j}=\frac{M\,(E^2_{j,\mathbf{k}}+\Delta_\mathrm{c}\Delta_\mathrm{f}+\xi_{\mathrm{c},\mathbf{k}}\xi_{\mathrm{f},\mathbf{k}}-M^2)}{2\,E_{j,\mathbf{k}}(E^2_{j,\mathbf{k}}-E^2_{\bar{j},\mathbf{k}})}\,, \nonumber
\end{align}	
\[ \bar{\lambda} =
\begin{cases}
\mathrm{c}       & \quad \text{if } \lambda=\mathrm{f}\\
\mathrm{f}       & \quad \text{if } \lambda=\mathrm{c}\\
\end{cases}
\]
\[ \bar{j} =
\begin{cases}
\mathrm{1}       & \quad \text{if } j=\mathrm{2}\\
\mathrm{2}       & \quad \text{if } j=\mathrm{1}\\
\end{cases}
\]
and excitation energies
\begin{align}
\label{eq:Eqple}
&\left(E_{(j=1,2),\mathbf{k}}\right)^2=\frac{1}{2}\left(\Gamma_\mathbf{k}\pm\sqrt{\Gamma^2_\mathbf{k}+\Omega_\mathbf{k}+\tilde{\Omega}_\mathbf{k}}\right)\,,\nonumber \\
&\Gamma_\mathbf{k}=2\,M^2+\Delta_\mathrm{c}^2+\Delta_\mathrm{f}^2+\xi_\mathrm{c}^2(\mathbf{k})+\xi_\mathrm{f}^2(\mathbf{k}+\mathbf{q})\,,\nonumber \\
&\Omega_\mathbf{k}=-4\left(\xi_\mathrm{c}^2(\mathbf{k})+\Delta_\mathrm{c}^2\right)\left(\xi_\mathrm{f}^2(\mathbf{k}+\mathbf{q})+\Delta_\mathrm{f}^2\right)\,, \nonumber \\
&\tilde{\Omega}_\mathbf{k}=8\,M^2\left(\Delta_\mathrm{c}\Delta_\mathrm{f}+\xi_\mathrm{c}(\mathbf{k})\xi_\mathrm{f}(\mathbf{k}+\mathbf{q})-\frac{M^2}{2}\right)\,.
\end{align}
To compute the equilibrium state, we solved the above  mean-field gap equations  self-consistently for given equilibrium chemical potential $\mu$ determined  by the level $x$ of Co doping.
While the above equations may also be used to describe a quasi--thermal photoinduced state characterized by time--dependent temperature and chemical potential, the main effects of interest here  come from 
non--thermal deviations 
from such  quasi--thermal state, which occur 
prior to thermalization of the photoexcited QPs
between the Fermi sea pockets.

\subsection*{Non-thermal dynamics}
To study the photo-excited 
incoherent  dynamics that govern our experiment, we first  introduce a basis of Bogoliubov QPs,
defined by the transformation
\begin{align}
\label{eq:trafo}
&c_{\mathbf{k},\uparrow}=u_{\mathbf{k}}\alpha^\dagger_\mathbf{k}-v_{\mathbf{k}}\beta_\mathbf{k}+\bar{u}_{\mathbf{k}}\gamma^\dagger_\mathbf{k}+\bar{v}_{\mathbf{k}}\delta_\mathbf{k}\,, \nonumber \\
&c^\dagger_{-\mathbf{k},\downarrow}=v_{\mathbf{k}}\alpha^\dagger_\mathbf{k}+u_{\mathbf{k}}\beta_\mathbf{k}+\bar{v}_{\mathbf{k}}\gamma^\dagger_\mathbf{k}-\bar{u}_{\mathbf{k}}\delta_\mathbf{k}\,, \nonumber \\
&f_{\mathbf{k}+\mathbf{q},\uparrow}=w_{\mathbf{k}}\alpha^\dagger_\mathbf{k}+x_{\mathbf{k}}\beta_\mathbf{k}+\bar{w}_{\mathbf{k}}\gamma^\dagger_\mathbf{k}-\bar{x}_{\mathbf{k}}\delta_\mathbf{k}\,, \nonumber \\
&f^\dagger_{-\mathbf{k}-\mathbf{q},\downarrow}=x_{\mathbf{k}}\alpha^\dagger_\mathbf{k}-w_{\mathbf{k}}\beta_\mathbf{k}+\bar{x}_{\mathbf{k}}\gamma^\dagger_\mathbf{k}+\bar{w}_{\mathbf{k}}\delta_\mathbf{k}\,.	
\end{align}
Here, $u_{\mathbf{k}}$, $v_{\mathbf{k}}$, $w_{\mathbf{k}}$, $x_{\mathbf{k}}$, $\bar{u}_{\mathbf{k}}$, $\bar{v}_{\mathbf{k}}$, $\bar{w}_{\mathbf{k}}$, and $\bar{x}_{\mathbf{k}}$ are coherence factors that depend on the instantaneous SC and SDW order parameters.  
We include for simplicity only a single momentum $\mathbf{q}$ in $H_\Delta$ and $H_\mathrm{m}$ as discussed above. Since we are interested in SDW excitonic state formation, we  consider the full inter--pocket SDW interaction Eq.~(\ref{eq:Hm}) 
without factorization, which introduces relaxation 
driven by the inter--band interaction. 
On the other hand, the SC interaction is treated  within the mean-field approximation for simplicity. 

Substituting Eq.(\ref{eq:trafo}) into the above Hamiltonian and eliminating the off--diagonal quadratic contributions,
we transform the Hamiltonian in the QP basis for given  
order parameters $\Delta_\mathrm{c}$, $\Delta_\mathrm{f}$ and $M$: 
\begin{align}
\label{eq:Hbcs_trafo}
H_\mathrm{BCS}&=H_0+H^\mathrm{MF}_\Delta \nonumber \\
&=\sum_\mathbf{k}\left[R^{-}_\mathbf{k}\left(\alpha^\dagger_\mathbf{k}\alpha_\mathbf{k}+\beta^\dagger_\mathbf{k}\beta_\mathbf{k}\right)+R^{+}_\mathbf{k}\left(\gamma^\dagger_\mathbf{k}\gamma_\mathbf{k}+\delta^\dagger_\mathbf{k}\delta_\mathbf{k}\right)\right],
\end{align}
where we introduced  
\begin{align}
&R^{-}_\mathbf{k}=\xi_\mathrm{c}(\mathbf{k})\left(v^2_{\mathbf{k}}-u^2_{\mathbf{k}}\right)+\xi_\mathrm{f}(\mathbf{k}+\mathbf{q})\left(x^2_{\mathbf{k}}-w^2_{\mathbf{k}}\right) \nonumber \\ &\qquad -2\left(u_{\mathbf{k}}v_{\mathbf{k}}\Delta_\mathrm{c}+x_{\mathbf{k}}w_{\mathbf{k}}\Delta_\mathrm{f}\right)\,,\nonumber \\
&R^{+}_\mathbf{k}=\xi_\mathrm{c}(\mathbf{k})\left(\bar{v}^2_{\mathbf{k}}-\bar{u}^2_{\mathbf{k}}\right)+\xi_\mathrm{f}(\mathbf{k}+\mathbf{q})\left(\bar{x}^2_{\mathbf{k}}-\bar{w}^2_{\mathbf{k}}\right) \nonumber \\ &\qquad -2\left(\bar{u}_{\mathbf{k}}\bar{v}_{\mathbf{k}}\Delta_\mathrm{c}+\bar{x}_{\mathbf{k}}\bar{w}_{\mathbf{k}}\Delta_\mathrm{f} \right)\,.
\end{align}	
and 
\begin{align}
\label{eq:Hm_trafo}
&H_\mathrm{m}=-\frac{V_\mathrm{m}}{2}\sum_{\mathbf{k},\mathbf{p}}\left[2\,l_{\mathbf{k}}m_{\mathbf{p}}\left(\alpha^\dagger_\mathbf{p}\alpha_\mathbf{p}+\beta^\dagger_\mathbf{p}\beta_\mathbf{p}\right)\right. \nonumber \\ &\left.+2\,l_{\mathbf{k}}p_{\mathbf{p}}\left(\gamma^\dagger_\mathbf{p}\gamma_\mathbf{p}+\delta^\dagger_\mathbf{p}\delta_\mathbf{p}\right)\right. \nonumber \\ &\left.
+m_{\mathbf{k}}m_{\mathbf{p}}\left(\alpha^\dagger_\mathbf{p}\alpha_\mathbf{p}+\beta^\dagger_\mathbf{p}\beta_\mathbf{p}\right)\left(\alpha^\dagger_\mathbf{k}\alpha_\mathbf{k}+\beta^\dagger_\mathbf{k}\beta_\mathbf{k}\right)
\right. \nonumber \\ &\left.
+p_{\mathbf{k}}p_{\mathbf{p}}\left(\gamma^\dagger_\mathbf{p}\gamma_\mathbf{p}+\delta^\dagger_\mathbf{p}\delta_\mathbf{p}\right)\left(\gamma^\dagger_\mathbf{k}\gamma_\mathbf{k}+\delta^\dagger_\mathbf{k}\delta_\mathbf{k}\right)
\right. \nonumber \\ &\left.
+p_{\mathbf{k}}m_{\mathbf{p}}\left(\alpha^\dagger_\mathbf{p}\alpha_\mathbf{p}+\beta^\dagger_\mathbf{p}\beta_\mathbf{p}\right)\left(\gamma^\dagger_\mathbf{k}\gamma_\mathbf{k}+\delta^\dagger_\mathbf{k}\delta_\mathbf{k}\right)
\right. \nonumber \\ &\left. +\left(r_{\mathbf{k}}r_{\mathbf{p}}+s_{\mathbf{k}}s_{\mathbf{p}}\right)\tilde{S}_z^\dagger(\mathbf{p})\tilde{S}(\mathbf{k})
\right. \nonumber \\ &\left.
+ \left(\bar{r}_{\mathbf{k}}\bar{r}_{\mathbf{p}}+\bar{s}_{\mathbf{k}}\bar{s}_{\mathbf{p}}\right)\bar{S}_z^\dagger(\mathbf{p})\bar{S}(\mathbf{k})
\right] 
\end{align}
with coherence factors
\begin{align}
&l_{\mathbf{k}}=2\left(u_\mathbf{k}w_\mathbf{k}+\bar{u}_\mathbf{k}\bar{w}_\mathbf{k}\right)\,,\quad m_{\mathbf{k}}=\left(u_\mathbf{k}w_\mathbf{k}+v_\mathbf{k}x_\mathbf{k}\right)\,,\nonumber \\ &p_{\mathbf{k}}=\left(\bar{u}_\mathbf{k}\bar{w}_\mathbf{k}+\bar{v}_\mathbf{k}\bar{x}_\mathbf{k}\right)\,, r_{\mathbf{k}}=v_\mathbf{k}\bar{w}_\mathbf{k}+\bar{v}_\mathbf{k}w_\mathbf{k}\,,\nonumber \\ &s_{\mathbf{k}}=\left(\bar{u}_\mathbf{k}x_\mathbf{k}+\bar{x}_\mathbf{k}u_\mathbf{k}\right)\,,\quad \bar{r}_\mathbf{k}=\bar{v}_\mathbf{k}x_\mathbf{k}+u_\mathbf{k}\bar{w}_\mathbf{k}\,, \nonumber \\ &\bar{s}_\mathbf{k}=v_\mathbf{k}\bar{x}_\mathbf{k}+\bar{u}_\mathbf{k}w_\mathbf{k}.
\end{align}
The last two lines in Equation (\ref{eq:Hm_trafo}) describe the
deviations from the mean field Hamiltonian and involve four QP operators (two  pairs of QPs). The  collective effects in the SDW channel are described by  the QP pair operators 
\begin{align}
\tilde{S}_z(\mathbf{k})=\beta_\mathbf{k}\gamma_\mathbf{k}+\alpha_\mathbf{k}\delta_\mathbf{k}\,, \quad \bar{S}_z(\mathbf{k})=\beta^\dagger_\mathbf{k}\delta_\mathbf{k}-\alpha^\dagger_\mathbf{k}\gamma_\mathbf{k}\,.
\end{align}
Since we are interested in long timescales 
after dephasing of any coherences among 
QPs, 
we only keep the QP number--conserving terms
in Eqs.~(\ref{eq:Hbcs_trafo}) and (\ref{eq:Hm_trafo}) and neglect any photoinduced 
coherence among QPs, 
$\langle\tilde{S}_z(\mathbf{k})\rangle=\langle\bar{S}_z(\mathbf{k})\rangle=0$.
The transformation of the SC and SDW gap equations in the QP basis then yields
\begin{align}
\label{eq:deltac}
&\Delta_\mathrm{c}=-V_\mathrm{SC}\sum_\mathbf{k}\left[u_{\mathbf{k}}v_{\mathbf{k}}\left(1-n^\alpha_\mathbf{k}-n^\beta_\mathbf{k}\right)\right. \nonumber \\ &\left.+\bar{u}_{\mathbf{k}}\bar{v}_{\mathbf{k}}\left(1-n^\gamma_\mathbf{k}-n^\delta_\mathbf{k}\right)\right]\,, \nonumber \\
&\Delta_\mathrm{f}=-V_\mathrm{SC}\sum_\mathbf{k}\left[w_{\mathbf{k}}x_{\mathbf{k}}\left(1-n^\alpha_\mathbf{k}-n^\beta_\mathbf{k}\right)\right. \nonumber \\ &\left.+\bar{w}_{\mathbf{k}}\bar{x}_{\mathbf{k}}\left(1-n^\gamma_\mathbf{k}-n^\delta_\mathbf{k}\right)\right]\,, \nonumber \\
&M=-\frac{V_\mathrm{m}}{2}\sum_\mathbf{k}\left[2\,l_{\mathbf{k}}-m_\mathbf{k}\left(n^\alpha_\mathbf{k}+n^\beta_\mathbf{k}\right)-p_\mathbf{k}\left(n^\gamma_\mathbf{k}+n^\delta_\mathbf{k}\right)\right]\,,
\end{align}
where we introduced the QP distributions
\begin{align}
&n^\alpha_\mathbf{k}=\langle\alpha^\dagger_\mathbf{k}\alpha_\mathbf{k}\rangle\,,\quad n^\beta_\mathbf{k}=\langle\beta^\dagger_\mathbf{k}\beta_\mathbf{k}\rangle\,,\quad n^\gamma_\mathbf{k}=\langle\gamma^\dagger_\mathbf{k}\gamma_\mathbf{k}\rangle\,, \nonumber \\ &n^\delta_\mathbf{k}=\langle\delta^\dagger_\mathbf{k}\delta_\mathbf{k}\rangle\,.
\end{align}	
As in the simple BCS theory, the excitation of QP populations quenches  both the SC and the SDW order parameters.  
We simplify the problem by assuming that, 
after the initial sub--ps QP relaxation, 
the QP distributions are all similar
for  high frequency optical pump  excitation at $\sim$1.5eV:  
$n_{\mathbf{k}}^\alpha \approx n_{\mathbf{k}}^\beta\equiv n^{\alpha\beta}_\mathbf{k}$ and $n_{\mathbf{k}}^\gamma \approx n_{\mathbf{k}}^\delta\equiv n^{\gamma\delta}_\mathbf{k}$.  
The time evolution of  the SC and SDW order parameters monitored by the THz probe 
is  determined  
by the time evolution of the above 
QP populations,
which we describe by deriving equations of motion 
using the full above Hamiltonian. 
The scattering processes 
determined by $H_m$ lead to QP relaxation  
described by 
\begin{align}
\label{eq:nk}
&\frac{\partial}{\partial t}n^{\alpha\beta}_\mathbf{p}=\frac{V_\mathrm{m}}{\hbar}\mathrm{Im}\sum_\mathbf{k}\left[\left(r_{\mathbf{k}} r_{\mathbf{p}}+s_{\mathbf{k}} s_{\mathbf{p}}\right)C_\mathrm{SDW,1}^{\mathbf{k},\mathbf{p}}\right.\nonumber \\ &\left.+\left(\bar{r}_{\mathbf{k}} \bar{r}_{\mathbf{p}}+\bar{s}_{\mathbf{k}} \bar{s}_{\mathbf{p}}\right)C_\mathrm{SDW,2}^{\mathbf{k},\mathbf{p}}\right]\,,\nonumber \\
&\frac{\partial}{\partial t}n^{\gamma\delta}_\mathbf{p}=\frac{V_\mathrm{m}}{\hbar}\mathrm{Im}\sum_\mathbf{k}\left[\left(r_{\mathbf{k}} r_{\mathbf{p}}+s_{\mathbf{k}} s_{\mathbf{p}}\right)C_\mathrm{SDW,1}^{\mathbf{k},\mathbf{p}}\right.\nonumber \\ &\left.-\left(\bar{r}_{\mathbf{k}} \bar{r}_{\mathbf{p}}+\bar{s}_{\mathbf{k}} \bar{s}_{\mathbf{p}}\right)C_\mathrm{SDW,2}^{\mathbf{k},\mathbf{p}}\right]\,.
\end{align}
The higher density matrices 
$C_{SDW}$ that appear on the right-hand side of the above 
equations 
involve four QP operators and 
are defined after subtracting all
factorizable contributions by using a cluster-expansion 
as in ~\cite{Book:11}:
\begin{align}
\label{eq:SDW_corr}
&C_{\mathrm{SDW,1}}^{\mathbf{p},\mathbf{k}}=\Delta\langle \tilde{S}^\dagger_z(\mathbf{k})\tilde{S}_z(\mathbf{p})\rangle\,, \quad C_{\mathrm{SDW,2}}^{\mathbf{p},\mathbf{k}}=\Delta\langle \bar{S}^\dagger_z(\mathbf{k})\bar{S}_z(\mathbf{p})\rangle\,.	
\end{align}	
For $C_{SDW}$=0 we recover the mean-field results, 
while the  four--QP density matrices  describe  correlation build--up and  scattering among QPs.  
Such processes modify the  QP distributions as compared to mean field, 
which  leads to time--dependent  SC order parameter quench.
The equations of motion of  $C_{SDW}$  describe the time evolution of correlations 
among the photoexcited QPs 
and 
are derived  
similar to Ref.~\cite{Book:11}:  
\begin{align}
\label{eq:dyn_corr1}
&\mathrm{i}\hbar\frac{\partial}{\partial t}C_\mathrm{SDW,1}^{\mathbf{p},\mathbf{k}}=\left(\varepsilon^{-}_\mathbf{p}+\varepsilon^+_\mathbf{p}-\varepsilon^{-}_\mathbf{k}-\varepsilon^+_\mathbf{k}\right)C_\mathrm{SDW,1}^{\mathbf{p},\mathbf{k}}+S_1^{\mathbf{p},\mathbf{k}} \nonumber \\
&+2\,V_\mathrm{m}\left(1-n^{\alpha\beta}_\mathbf{k}-n^{\gamma\delta}_\mathbf{k}\right)\sum_\mathbf{l}\left(r_{\mathbf{k}}r_{\mathbf{l}}+s_{\mathbf{k}}s_{\mathbf{l}}\right)C_\mathrm{SDW,1}^{\mathbf{p},\mathbf{l}}\nonumber \\
&-2\,V_\mathrm{m}\left(1-n^{\alpha\beta}_\mathbf{k}-n^{\gamma\delta}_\mathbf{k}\right)\sum_\mathbf{l}\left(r_{\mathbf{p}}r_{\mathbf{l}}+s_{\mathbf{p}}s_{\mathbf{l}}\right)C_\mathrm{SDW,1}^{\mathbf{l},\mathbf{k}}\nonumber \\&+D_1^{\mathbf{p},\mathbf{k}}+T_1^{\mathbf{p},\mathbf{k}}\,,
\end{align}
\begin{align}	
\label{eq:dyn_corr2}
&\mathrm{i}\hbar\frac{\partial}{\partial t}C_\mathrm{SDW,2}^{\mathbf{p},\mathbf{k}}=\left(\varepsilon^{-}_\mathbf{p}-\varepsilon^+_\mathbf{p}-\varepsilon^{-}_\mathbf{k}+\varepsilon^+_\mathbf{k}\right)C_\mathrm{SDW,2}^{\mathbf{p},\mathbf{k}}+S_2^{\mathbf{p},\mathbf{k}} \nonumber \\
&+2\,V_\mathrm{m}\left(n^{\alpha\beta}_\mathbf{k}-n^{\gamma\delta}_\mathbf{k}\right)\sum_\mathbf{l}\left(\bar{r}_{\mathbf{k}}\bar{r}_{\mathbf{l}}+\bar{s}_{\mathbf{k}}\bar{s}_{\mathbf{l}}\right)C_\mathrm{SDW,2}^{\mathbf{p},\mathbf{l}}\nonumber \\
&-2\,V_\mathrm{m}\left(n^{\alpha\beta}_\mathbf{k}-n^{\gamma\delta}_\mathbf{k}\right)\sum_\mathbf{l}\left(\bar{r}_{\mathbf{p}}\bar{r}_{\mathbf{l}}+\bar{s}_{\mathbf{p}}\bar{s}_{\mathbf{l}}\right)C_\mathrm{SDW,2}^{\mathbf{l},\mathbf{k}}\nonumber \\&+D_2^{\mathbf{p},\mathbf{k}}+T_2^{\mathbf{p},\mathbf{k}}\,, 	
\end{align}
where the  QP  energies
are 
given by 
\begin{align}
\label{eq:Wannier_Eqple}
&\varepsilon^{-}_\mathbf{k}=\xi_\mathrm{c}(\mathbf{k})\left(v^2_\mathbf{k}-u^2_\mathbf{k}\right)+\xi_\mathrm{f}(\mathbf{k}+\mathbf{q})\left(x^2_\mathbf{k}-w^2_\mathbf{k}\right)\nonumber \\&-2\left(u_\mathbf{k}v_\mathbf{k}\Delta_\mathrm{c}+x_\mathbf{k}w_\mathbf{k}\Delta_\mathrm{f}\right)-2\left(u_\mathbf{k}w_\mathbf{k}+v_\mathbf{k}x_\mathbf{k}\right)M\,,\nonumber \\
&\varepsilon^{+}_\mathbf{k}=\xi_\mathrm{c}(\mathbf{k})\left(\bar{v}^2_\mathbf{k}-\bar{u}^2_\mathbf{k}\right)+\xi_\mathrm{f}(\mathbf{k}+\mathbf{q})\left(\bar{x}^2_\mathbf{k}-\bar{w}^2_\mathbf{k}\right)\nonumber \\ &-2\left(\bar{u}_\mathbf{k}\bar{v}_\mathbf{k}\Delta_\mathrm{c}+\bar{x}_\mathbf{k}\bar{w}_\mathbf{k}\Delta_\mathrm{f}
\right)-2\left(\bar{u}_\mathbf{k}\bar{w}_\mathbf{k}+\bar{v}_\mathbf{k}\bar{x}_\mathbf{k}\right)M \,.
\end{align}
The usual scattering 
among individual QPs 
is described 
by the  source terms
\begin{align}
\label{S1}
&S_1^{\mathbf{p},\mathbf{k}}=\frac{4\,V_\mathrm{m}}{S}\left(r_{\mathbf{k}}r_{\mathbf{p}}+s_{\mathbf{k}}s_{\mathbf{p}}\right)\left[n^{\alpha\beta}_\mathbf{p}n^{\gamma\delta}_\mathbf{p}(1-n^{\alpha\beta}_\mathbf{k})(1-n^{\gamma\delta}_\mathbf{k})\right.\nonumber \\ &\left.-n^{\alpha\beta}_\mathbf{k}n^{\gamma\delta}_\mathbf{k}(1-n^{\alpha\beta}_\mathbf{p})(1-n^{\gamma\delta}_\mathbf{p})\right]\,, 
\end{align}
\begin{align}
\label{S2}
&S_2^{\mathbf{p},\mathbf{k}}=\frac{4\,V_\mathrm{m}}{S}\left(\bar{r}_{\mathbf{k}}\bar{r}_{\mathbf{p}}+\bar{s}_{\mathbf{k}}\bar{s}_{\mathbf{p}}\right)\left[n^{\alpha\beta}_\mathbf{k}n^{\gamma\delta}_\mathbf{p}(1-n^{\alpha\beta}_\mathbf{p})(1-n^{\gamma\delta}_\mathbf{k})\right.\nonumber \\ &\left.-n^{\alpha\beta}_\mathbf{k}n^{\gamma\delta}_\mathbf{p}(1-n^{\alpha\beta}_\mathbf{p})(1-n^{\gamma\delta}_\mathbf{k})\right]\,.
\end{align}
$S_{1,2}^{\mathbf{p},\mathbf{k}}$ have the typical form describing Boltzmann scattering
with in- and out-scattering contributions.
The  first line in  
Eqs.~(\ref{eq:dyn_corr1}) and (\ref{eq:dyn_corr2})
describe relaxation  among individual quasi--particles 
within the Born approximation, 
without any excitonic correlation. 
Such perturbative Born scattering approximation does not change  the behavior at long 100's ps times.   

The next two lines on the rhs of  Eqs.~(\ref{eq:dyn_corr1}) and (\ref{eq:dyn_corr2})
give the most important  
contributions here. As soon as  
non--thermal QP populations are excited in the $e$ and $h$ Fermi sea pockets,  
$S^{\mathbf{p},\mathbf{k}} \ne $0 and the above equations  describe  screening build--up and formation of 
spin--excitons
among the laser-induced QPs.
Renormalization of the QP energies and screening-type effects are described by the remaining two-particle contributions  $D_{1,2}^{\mathbf{p},\mathbf{k}}$.
The coupling to  three-particle correlations, $T_{1,2}^{\mathbf{p},\mathbf{k}}$, introduces  relaxation of the SDW excitonic correlation.

$C^{\mathbf{p},\mathbf{k}}_\mathrm{SDW,2}$ is mostly significant   in the strong excitation regime, 
as it requires an appreciable imbalance between QP distributions 
such that $(n^{\alpha\beta}_\mathbf{k}-n^{\gamma\delta}_\mathbf{k})$ is non-vanishing. 
In contrast, $C^{\mathbf{p},\mathbf{k}}_\mathrm{SDW,1}$ becomes large already at low QP densities. Here 
we  assume 
that high--frequency pump excitation results in similar 
nonthermal densities of $\alpha$-, $\beta$- and $\gamma$-, $\delta$- QPs, so
we  neglect $C^{\mathbf{p},\mathbf{k}}_\mathrm{SDW,2}$.
More details on the full theory will be presented elsewhere \cite{theo_paper}. 

\subsection*{Generalized Wannier equation for describing the excitonic correlation}
Following an initial 
temporal regime of ultrafast SC gap quenching,
the QP distributions $n^{\alpha\beta}_\mathbf{k}$ and $n^{\gamma\delta}_\mathbf{k}$ change 
adiabatically with time, so we  seek stationary  solutions of Eq.~(\ref{eq:dyn_corr1}).
The form of these equations of motion 
suggests 
the transformation of  the SDW correlation $C_{\mathrm{SDW,1}}^{\mathbf{p},\mathbf{k}}$ into an excitonic  basis \cite{Book:11}
defined by the wavefunction $\phi^\mathrm{r}_{\nu,\mathbf{q}}$. The latter is 
given 
by the generalized Wannier equation 
\begin{align}
\label{eq:Wannier}	&\left(\varepsilon^{-}_\mathbf{p}+\varepsilon^{+}_\mathbf{p}\right)\phi^\mathrm{r}_\nu(\mathbf{p})-\left(1-n^{\alpha\beta}_\mathbf{p}-n^{\gamma\delta}_\mathbf{p}\right)\sum_\mathbf{k}V_{\mathbf{k},\mathbf{p}}\phi^\mathrm{r}_\nu(\mathbf{k})\nonumber \\ &=E_\nu\,\phi^\mathrm{r}_\nu(\mathbf{p})\,,
\end{align}
where the  Coulomb matrix element depends on the QP coherence factors:
\begin{align}	
V_{\mathbf{k},\mathbf{p}}=2\,V_\mathrm{m}&\left[\left(\bar{u}_\mathbf{k}x_\mathbf{k}+\bar{x}_\mathbf{k}u_\mathbf{k}\right)\left(\bar{u}_\mathbf{p}x_\mathbf{p}+\bar{x}_\mathbf{p}u_\mathbf{p}\right)\right. \nonumber \\ &\left.+\left(v_\mathbf{k}\bar{w}_\mathbf{k}+\bar{v}_\mathbf{k}w_\mathbf{k}\right)\left(v_\mathbf{p}\bar{w}_\mathbf{p}+\bar{v}_\mathbf{p}w_\mathbf{p}\right)\right]\,.
\end{align}
It is then convenient to introduce the excitonic  operator
\begin{align}	X_{\nu}=\sum_\mathbf{p} \phi^\mathrm{l *}_{\nu}(\mathbf{p}) \tilde{S}_z(\mathbf{p}).
\,
\end{align}
Unlike for phonons, the commutation relations of this  composite exciton operator 
have non-bosonic corrections 
due to Phase Space Filling arising from the fermionic character of the QPs involved.
By transforming from uncorrelated QPs
to the excitonic basis 
\begin{align}
\tilde{S}_z(\mathbf{p})=\sum_\nu\phi^\mathrm{r}_{\nu}(\mathbf{p})X_{\nu}\,
\end{align}
we describe the correlations of interest in terms of the above-defined  spin-excitons: 
\begin{align}
\label{eq:corr_trafo}
&C_\mathrm{SDW,1}^{\mathbf{p},\mathbf{k}}=\sum_{\nu,\nu'}\left[\phi^\mathrm{r}_{\nu}(\mathbf{k})\right]^\star\phi^\mathrm{r}_{\nu'}(\mathbf{p})\,\Delta\langle X^\dagger_{\nu}X_{\nu'}\rangle\,,\nonumber \\
&\Delta\langle X^\dagger_{\nu}X_{\nu'}\rangle=\sum_{\mathbf{k},\mathbf{p}}\phi^\mathrm{l}_{\nu}(\mathbf{k})\left[\phi^\mathrm{l}_{\nu'}(\mathbf{p})\right]^\star C_\mathrm{SDW,1}^{\mathbf{p},\mathbf{k}}\,.
\end{align}	
The coupling of the QP distributions 
to the excitonic amplitude 
in Eq.~(\ref{eq:Wannier}) yields a non-hermitian eigenvalue problem, so we obtain  left- and right-handed eigenfunctions $\phi^\mathrm{r,l}_{\nu,\mathbf{q}}$. 
These 
describe both bound and scattering solutions, 
where the latter correspond to unbound QP pairs
whose properties are modified by the magnetic interaction.
The above wavefunctions  satisfy the orthogonality and completeness relations
\begin{align}
\sum_\mathbf{p}\left[\phi^\mathrm{l}_{\nu}(\mathbf{p})\right]^\star\phi^\mathrm{r}_{\nu'}(\mathbf{p})=\delta_{\nu,\nu'}\,,\quad \sum_\nu\left[\phi^\mathrm{l}_{\nu}(\mathbf{p})\right]^\star\phi^\mathrm{r}_{\nu}(\mathbf{p}')=\delta_{\mathbf{p},\mathbf{p}'}\,.
\end{align}	
To simplify the problem for our purposes here, we 
assume relaxation to 
the lowest spin--exciton state  $\phi_\mathbf{p}\equiv\phi^\mathrm{r}_{\nu=0}(\mathbf{p})$ and only retain
this contribution to Eq.~(\ref{eq:corr_trafo}). As a result, 
\begin{align}
\label{eq:corr}
C^{\mathbf{p},\mathbf{k}}_\mathrm{SDW,1}=\phi^\star_\mathbf{p}\phi_\mathbf{k}\,,
\end{align}
where we have absorbed $\Delta\langle X^\dagger_{\nu=0} X_{\nu=0}\rangle$ into $\phi_\mathbf{p}$.

In the incoherent long--time regime and for $C^{\mathbf{p},\mathbf{k}}_\mathrm{SDW,1}$ dominating over $C^{\mathbf{p},\mathbf{k}}_\mathrm{SDW,2}$, 
we obtain  Eq.~(\ref{eq:conserv})
from an exact relation between the traces of the corresponding density matrices. \cite{theo_paper}.
The coupled Eqs.~(\ref{eq:conserv}) and (\ref{eq:Wannier}),  together with the order parameter equations (\ref{eq:deltac}) and coherent factor expressions, yield a self-consistent 
calculation of the 
many-body state 
defined by  $(\phi_\mathbf{p},n_\mathbf{p},\Delta_\mathrm{c},\Delta_\mathrm{f},M)$. 
This result corresponds to an adiabatic  solution of the equations of motion 
and describes the non-equilibrium state reached after formation/buildup of SDW correlation 
and before the system thermalizes  via  scattering  across the Fermi sea pockets. 

\subsection*{Numerical Calculations}
In our numerical calculations presented in Fig. 4, 
we first computed the thermal 
ground state configuration by solving the SDW and SC gap equations~(\ref{eq:gap_gs}) self-consistently for a given doping level. We then solved Eqs.  (\ref{eq:Wannier0}) and ~(\ref{eq:conserv}) together with the order parameter  equations  iteratively until convergence was reached. The energy eigenvalue $E$ determines the total  QP density $\rho=1/S\sum_\mathbf{k}n_\mathbf{k}$
and thus corresponds 
to  excitonic corrections to the chemical potential. 
In all numerical calculations we used typical parameters of $\mathrm{Ba}(\mathrm{Fe}_{1-x}\mathrm{Co}_x)_2\mathrm{As}_2$, which yield a good qualitative agreement with the experimentally observed doping dependence of the SC and SDW orders~\cite{Fernandes:2010}.
To model the photoinduced initial condition  immediately after the initial 
phonon--induced ultrafast SC gap quench 
following photocarrier relaxation, we assume that the  excited QPs have relaxed  into the different pockets 
and  describe their distributions for simplicity  by Fermi--Dirac distributions 
\begin{align}
\label{eq:FD}
&n^{\alpha,\beta}_\mathbf{k}=\frac{1}{1+\mathrm{exp}\left(\varepsilon^{-}_\mathbf{k}/k_\mathrm{B}T_{\alpha,\beta}\right)}\,, \nonumber \\ &n^{\gamma,\delta}_\mathbf{k}=\frac{1}{1+\mathrm{exp}\left(\varepsilon^{+}_\mathbf{k}/k_\mathrm{B}T_{\gamma,\delta}\right)}\,
\end{align}
which determine the initial condition to our time--dependent calculation. 
The temperatures $T_{\alpha,\beta}$ and $T_{\gamma,\delta}$ 
are obtained from the total QP densities
\begin{align}
\rho_{\alpha,\beta}=\frac{1}{S}\sum_\mathbf{k}n_\mathbf{k}^{\alpha,\beta}\,,\quad \rho_{\gamma,\delta}=\frac{1}{S}\sum_\mathbf{k}n_\mathbf{k}^{\gamma,\delta}\,.
\end{align}
In the actual calculations presented here, $T_{\alpha,\beta}$ and $T_{\gamma,\delta}$ were chosen such that the total densities of the different QPs are the same, i.~e. $\rho\equiv \rho_{\alpha,\beta}= \rho_{\gamma,\delta}$.
However, our conclusions do not depend on 
how we describe the initial QP distributions, 
which  form following relaxation
of the photocarriers from  high energy energy states populated by the pump that are not well known. 

\end{document}